\documentclass[twocolumn,prl]{revtex4}
\usepackage{graphics,graphicx}
\usepackage{ulem}
\begin{document}
\title{Magnetic fluctuations driven insulator-to-metal transition in Ca(Ir$_{1-x}$Ru$_{x}$)O$_{3}$}
\author{J.~Gunasekera$^{1}$}
\author{L. ~Harriger$^{2}$}
\author{A. ~Dahal$^{1}$}
\author{T. ~Heitmann$^{3}$}
\author{G. ~Vignale$^{1}$}
\author{D. K.~Singh$^{1,*}$}
\affiliation{$^{1}$Department of Physics and Astronomy, University of Missouri, Columbia, MO 65211}
\affiliation{$^{2}$National Institute of Standards and Technology, Gaithersburg, MD 20899}
\affiliation{$^{3}$University of Missouri Research Reactor, University of Missouri, Columbia, MO 65211}
\affiliation{$^{*}$Email: singhdk@missouri.edu}

\begin{abstract}

{\bf Magnetic fluctuations in transition metal oxides are a subject of intensive research because of the key role they are expected to play in the transition from the Mott insulator to the unconventional metallic phase of these materials, and also as drivers of superconductivity. Despite much effort, a clear link between magnetic fluctuations and the insulator-to-metal transition has not yet been established.  Here we report the discovery of a compelling link between magnetic fluctuations and the insulator-to-metal transition in Ca(Ir$_{1-x}$Ru$_{x}$)O$_{3}$ perovskites as a function of the doping coefficient x. We show that when the material turns from insulator to metal, at a critical value of x$\sim$ 0.3, magnetic fluctuations change their character from antiferromagnetic, a Mott insulator phase,  to ferromagnetic, an itinerant electron state with Hund's orbital coupling. These results are expected to have wide-ranging implications for our understanding of the unconventional properties of strongly correlated electrons systems.}

\end{abstract}

 \maketitle

 The $d$-band metal oxides are most notorious among materials for the complexity of their phase diagram, displaying a formidable array of close and often overlapping transitions between metallic, insulating, magnetic, and even superconducting phases.\cite{Varma,Emery}  After decades of intensive study a full theoretical understanding of this perplexing phase diagram remains elusive.\cite{Tranquada,McLaughlin,Fisk,Pengcheng,Hwang,Tokura}  In a typical scenario, one starts from an insulating, antiferromagnetically ordered state when the d-band is half-filled or nearly so.  This is a classic example of a Mott insulator in which the single occupancy of each lattice site prevents free motion of the charge.\cite{Verwey,Mott,Mott2}  Moving away from the half-filled situation (by chemical doping, for example) the charge eventually unfreezes, leading to a metallic phase with striking non Fermi liquid properties, and in some cases to a superconducting phase.\cite{Anderson,Hubbard,Kanamori}   In parallel to this, the nature of magnetic correlation changes dramatically from predominantly antiferromagnetic (superexchange-like)  in the Mott-localized  phase to predominantly ferromagnetic (Hund-like) in the delocalized metallic phase. Even in the absence of the long-range magnetic order, the nature of the magnetic fluctuations is of paramount importance to superconductivity. Since these fluctuations are believed to be the mediators of the attractive interaction that leads to the formation of Cooper pairs,\cite{Imada,Okamoto,Jackeli,Qazilbash,Hwang} the nature of magnetic fluctuations thus control the symmetry of the superconducting order parameter.\cite{Tokura,Lee,Orenstein}  

\begin{figure*}
\centering
\includegraphics[width=17 cm]{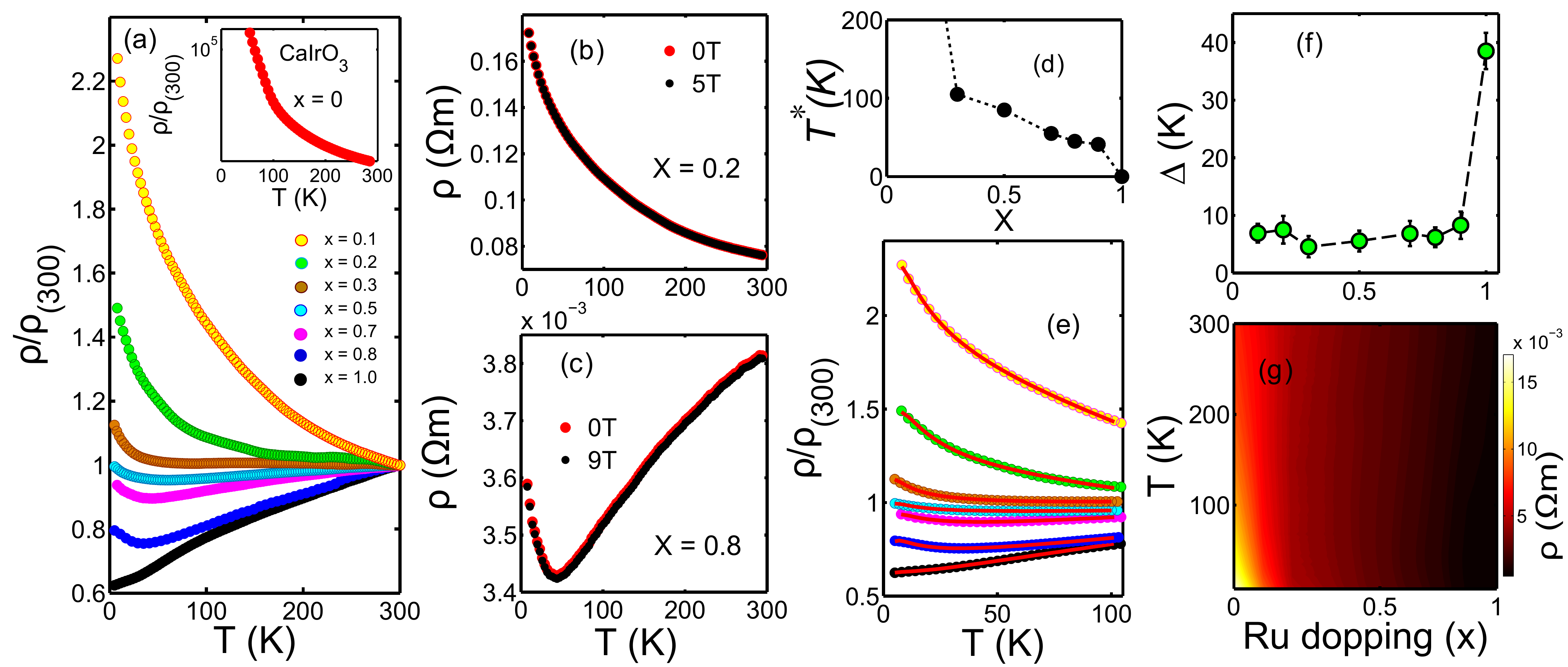} \vspace{-4mm}
\caption{\textbf{Insulator-to-metal transition in Ca(Ir$_{1-x}$Ru$_{x}$)O$_{3}$}. (a) Detailed electrical measurements exhibiting the insulator-to-metal transition as a function of x in Ca(Ir$_{1-x}$Ru$_{x}$)O$_{3}$. At x = 0, the compound CaIrO$_{3}$ is a strong Mott insulator (see inset). As x increases above 0.3, metallic characteristic becomes evident at high temperatures. At x = 1, the compound CaRuO$_{3}$ is a non-Fermi liquid metal. (b)-(c) Magnetic field has no effect on electrical properties in any composition. For illustration purposes, electrical resistivity data in applied fields are plotted for two chemical compositions. (d) Characteristic temperature $T$$^{*}$, indicating the onset of an upturn in resistivity in a given composition, as a function of x. (e) The electrical resistivity data below $T$$\leq$ 100 K are fitted with a model to extract preliminary information about the role of magnetism in the insulator-to-metal transition in the system. (f) The fitting parameter $\Delta$ increases significantly as x$\rightarrow$ 1 (see text for detail). (g) The insulator-to-metal transition process is summarized in a contour plot of resistivity.
} \vspace{-4mm}
\end{figure*}

Motivated by the above considerations, we have undertaken a careful experimental study of the interplay between the insulator to metal transition and magnetism in the perovskites Ca(Ir$_{1-x}$Ru$_{x}$)O$_{3}$. These compounds crystallize in an orthorhombic lattice configuration with end members, CaIrO$_{3}$ (x = 0) and CaRuO$_{3}$ (x = 1), being a Mott insulator and an anomalous metal, respectively.\cite{Ohgushi1,Cao2} It is well established that CaIrO$_{3}$ exhibits antiferromagnetic order for $T\leq110$ K.\cite{Sala2,Singh2} The nature of the magnetism in CaRuO$_{3}$  -- whether it is a paramagnet or on the verge of the ferromagnetic instability -- is still a matter of debate.\cite{Mazin,Cao} Recent experimental studies suggest a coexistence of the quantum magnetic fluctuations and non-Fermi liquid behavior for $T\leq 25$ K, albeit no magnetic order is detected to the lowest measurement temperature.\cite{Singh1} On the other hand, significantly different electric and magnetic properties are observed in CaIrO$_{3}$: sharp peaks in the static and the dynamic magnetic susceptibilities, indicating the absence of magnetic fluctuations, coincide with an upturn in the electrical resistivity at $T\simeq 110$ K.\cite{Singh2} Because the two end members of the group have so diverse physical and magnetic properties, we find them very suitable for a comprehensive investigation of the insulator-to-metal transition as a function of the chemical doping coefficient x. Experimentally, we have been able to pinpoint the insulator to metal transition at the relatively small value of the chemical doping, x=0.3. The concomitant change in the character of the magnetic fluctuations from static antiferromagnetic to dynamical ferromagnetic is clearly demonstrated in our data, presented below. Remarkably, the value x=0.3 coincides with the level of hole doping for which the wave vector dependent magnetic susceptibility of the Hubbard model changes from being predominantly antiferromagnetic (peak at wave vector ${\bf q}=(\pi,\pi)$) to be predominantly ferromagnetic (peak at ${\bf q}=0$).  Because of their accuracy and consistency these results provide a strong motivation for more comprehensive theories involving the interplay between spin fluctuations and the insulator-to-metal transition (see Fig.~\ref{PhaseDiagram} below). 

\textbf{Insulator-to-metal transition in perovskites Ca(Ir$_{1-x}$Ru$_{x}$)O$_{3}$}. We have performed detailed study of Ca(Ir$_{1-x}$Ru$_{x}$)O$_{3}$ compounds to elucidate the correlation between magnetic and electrical properties by systematically varying the chemical doping coefficient x in small steps of 0.1. The high quality polycrystalline samples were synthesized using solid-state reactions and characterized using a high-resolution powder X-ray diffractometer (see Methods and Supplementary Materials for detail). While the chemical substitution of Ir by Ru introduces small changes in the lattice parameters (see Fig. S1 of the Supplementary Materials), the orthorhombic crystal structure is preserved throughout the group with volume of a lattice unit cell at 229$\pm$2 $\AA$$^{3}$. 

\begin{figure}
\centering
\includegraphics[width=8.9 cm]{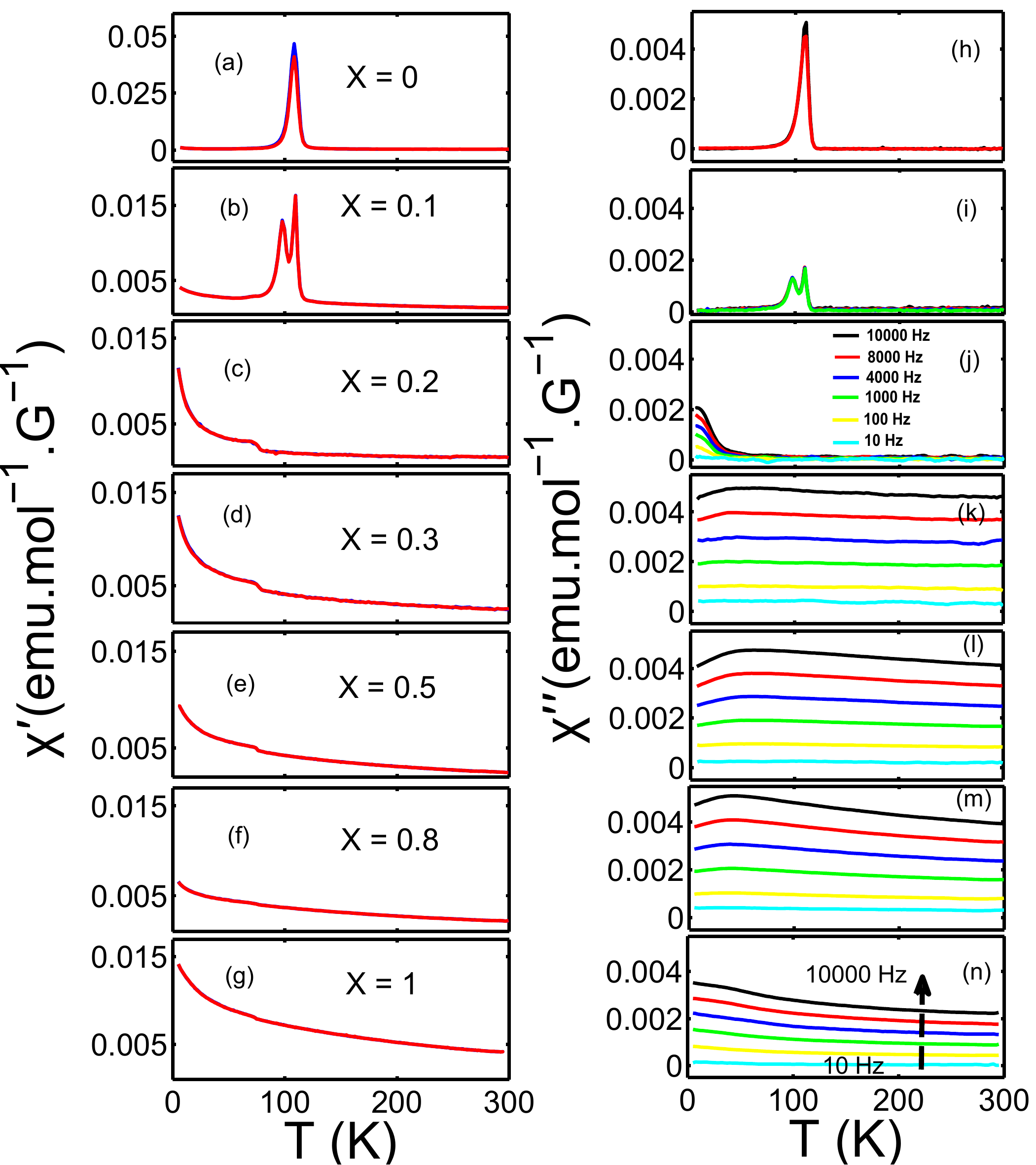} \vspace{-4mm}
\caption{\textbf{Evidence of magnetic fluctuations via detailed ac susceptibility measurements}. (a)-(g) Static susceptibilities ($\chi'$) at different frequencies (10 - 10$^{4}$ Hz) are plotted as a function of temperature for various chemical doping coefficients x in Ca(Ir$_{1-x}$Ru$_{x}$)O$_{3}$. A sharp peak, indicating the onset of long-range magnetic order at $T \simeq 110$ K in x = 0, gradually disappears and is replaced by a continuous increase in $\chi'$ at higher $x$ at low temperature. While the origin of the increase in $\chi'$ is still not clear, no magnetic order is detected in elastic neutron scattering measurements in x = 0.8 and x = 1 compositions (see supplementary materials for detail). (h)-(n) Dynamic susceptibilities ($\chi''$) at different frequencies are plotted as a function of temperature for various chemical doping coefficients x. Unlike $\chi'$, $\chi''$ exhibits strong frequency dependence in higher x compositions (x$\geq$ 0.3). The strong frequency dependence hints of the development of magnetic fluctuations in respective chemical dopings. Thus, long range magnetic order at low x (insulating phase) is replaced by magnetic fluctuations in higher x compositions (metallic phase).}
\end{figure}

The insulator-to-metal transition as a function of the coefficient x is evident from the electrical transport measurements, as shown in Fig. 1a. The normalized electrical resistivity ($\rho$/$\rho_{300}$, where $\rho_{300}$ is the resistivity at $T$ = 300 K) is plotted as a function of temperature for various chemical doping percentages. A gradual evolution from a strong insulating state, x = 0, to a completely metallic state is observed as the coefficient x varies from 0 to 1. An upturn in the electrical resistivity at low temperature is found to be present until x=0.9, albeit very weakly. This behavior is further characterized by a characteristic temperature, $T^*$, for a given coefficient x. The characteristic temperature $T^*$ decreases almost monotonically for x$\geq$0.3 and approaches $T^*$$\rightarrow$0 K as x$\rightarrow$ 1 (see Fig. 1d). At x = 1, the compound CaRuO$_{3}$ exhibits an anomalous metallic behavior at low temperature. Interestingly,  the application of a magnetic field has no pronounced effect on the electrical transport data in any of the doping percentages. Characteristic plots of $\rho$ vs. $T$ in applied fields are shown in Fig. 1b and 1c for two chemical compositions of x = 0.2 and 0.8, respectively. Electrical resistivity data are further analyzed using an expression that is often used to describe the electrical properties of underdoped cuprates and the spin ladder compounds.\cite{Moshchalkov} The expression essentially uses a single fitting parameter to correlate electrical behavior to the underlying magnetism.\cite{Moshchalkov} It is given by:

\begin{eqnarray*}
{\frac{\rho(T)}{\rho(300)}}&\propto& {\rho_0}+{a}{T}+{\frac{b}{T}}e^{-2\Delta/T}\,,
\end{eqnarray*}
where $\rho$$_{0}$ is the residual resistivity and $\Delta$ is the fitting parameter related to the spin or the charge gap in the system. If $\Delta>0$, this expression tends to a finite value for $T \to 0$, i.e., it describes a metal. For $\Delta<0$, the resistivity diverges at low $T$ (as found in CaIrO$_{3}$). In Fig. 1e, we have fitted the resistivity data of Ca(Ir$_{1-x}$Ru$_{x}$)O$_{3}$ below $T$$\leq$100 K using above expression. The parameter $\Delta$ reduces significantly from 39.6 K to 8.2 K as x varies from 1 to 0.1 (Fig. 1f). At x=0, $\Delta$ is found to be negative (-532 K). Since the perovskite Ca(Ir$_{1-x}$Ru$_{x}$)O$_{3}$ becomes fully metallic as x$\rightarrow$1, it is no surprise that CaRuO$_{3}$ (x=1) exhibits the largest positive value of $\Delta$. The transition to the metallic state as functions of the chemical doping coefficient x and temperature $T$ is summarized in a contour plot of resistivity in Fig. 1g.

\begin{figure*}
\centering
\includegraphics[width=16 cm]{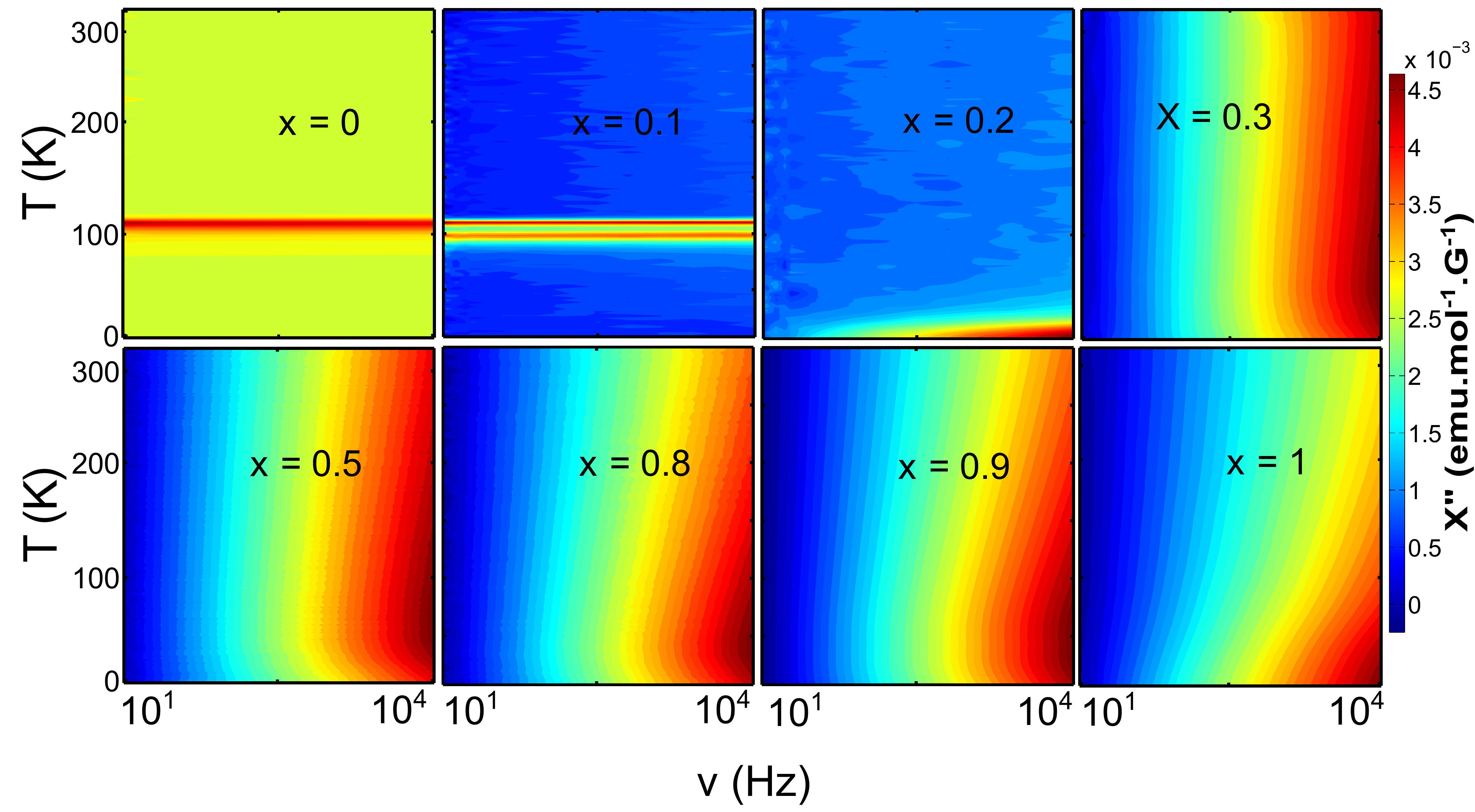} \vspace{-4mm}
\caption{\textbf{Contour plots of dynamic susceptibilities for various chemical doping percentages x in Ca(Ir$_{1-x}$Ru$_{x}$)O$_{3}$}. Temperature vs. frequency plots of dynamic susceptibilities outline the development of magnetic fluctuations as a function of the chemical doping coefficient x. $\chi$$^{"}$ is independent of ac frequency near x = 0. As the coefficient x increases, $\chi$$^{"}$ starts developing frequency-dependence. For x$\geq$0.3, strong frequency-dependence of the dynamic susceptibilities are clearly observable, indicating significant magnetic fluctuations at the long time scale of $t$$\geq$100 micro-second.  
} \vspace{-4mm}
\end{figure*}

\textbf{Magnetic fluctuations accompanying the insulator-to-metal transition}. Next, we investigate the role of magnetic fluctuations in the insulator to metal transition in Ca(Ir$_{1-x}$Ru$_{x}$)O$_{3}$. The ac susceptibility measurements provide information about the static and the dynamic magnetic properties of correlated spins at the same time.  The full ac susceptibility $\chi$ ($T$) is written as $\chi$($T$) = $\chi$$^{'}$ ($T$) + $i$ $\chi$$^{"}$ ($T$), where the real part $\chi$$^{'}$ represents the static magnetic behavior and the imaginary $\chi$$^{"}$ provides information about the dynamic magnetic properties or, the magnetic fluctuations in a system. The susceptibility measurements in applied ac frequencies of 10-10$^{4}$ Hz were performed for all chemical doping x. Experimental results are plotted in Fig. 2a-2n. The static and dynamic susceptibilities are found to be characteristically different in perovskites with coefficient x $\geq$~0.2, compared to x~$<$~0.2. In CaIrO$_{3}$ (x = 0), both $\chi$$^{'}$ and $\chi$$^{"}$ exhibits frequency-independent sharp peaks, suggesting a staggered antiferromagnetic configuration, setting in at a temperature $T$$\simeq$110 K.\cite{Singh2,Sala2,Ohgushi2} The absence of the frequency dependence of $\chi$$^{"}$ rules out any spin dynamics as functions of energy (frequency) and temperature. As x is varied from 0 to 0.1, single sharp peaks in $\chi$$^{'}$ and $\chi$$^{"}$ vs temperature are replaced by two-peaked structures. For further increase in x, the dynamic susceptibility is found to develop frequency dependence that becomes stronger as x increases. For x $\geq$0.3, the frequency-dependence of $\chi''$ extends up to $T$=300 K and develops a broad maximum in temperature at low temperature. This observation is in stark contrast to the frequency-independent character of $\chi'$. A noticeable feature in the static susceptibility involves the observation of a sharp cusp at $T\simeq 80$ K followed by a significant increase in $\chi'$ in x$\geq$0.2 compositions. While it is not clear whether the cusp is associated with the onset of a charge order or with a structural transition in the system, no magnetic order was detected in the detailed neutron scattering measurements on two chemical compositions of x = 1 and x = 0.8 (see Fig. S2 in Supplementary Materials). A comparison of the static and the dynamic susceptibilities in a given chemical composition shows that the system manifests significantly strong dynamic magnetic behavior in compounds with x$\geq$0.3.

The contour plots of the dynamic susceptibilities in Fig. 3 comprehensively illustrate the evolution of magnetic fluctuations, with long relaxation time, in Ca(Ir$_{1-x}$Ru$_{x}$)O$_{3}$ as a function of the chemical doping coefficient x. At x = 0, the plot is characterized by a singular line at $T$$\simeq$110 K, indicating the absence of dynamic behavior. As x increases above x$\geq$0.3, the small regime of large $\chi''$ at high frequency and low temperature (in x = 0.2) extends to higher temperature, thus illustrating significant spin dynamics in compounds with coefficient x$\geq$0.3. In fact, the dynamic response to the ac susceptibility measurements is found to be strongest in the x = 0.3 composition. A comparison of the contour plots in Fig. 3 to the electrical transport measurements in Fig. 1a reveals one-to-one correspondence between the development of magnetic fluctuation and the onset of the metallic behavior. At x = 0, the compound CaIrO$_{3}$ is a strong antiferromagnetic Mott insulator with no obvious dynamic properties at long time scale.\cite{Singh2,Bogdanov,Ohgushi2} At x = 1, the compound CaRuO$_{3}$ is a non-Fermi liquid metal with strong dynamic magnetic properties.\cite{Lee2,Klein,Singh1} Using this method, we have investigated magnetic fluctuations in various compositions of Ca(Ir$_{1-x}$Ru$_{x}$)O$_{3}$ perovskites on a time scale of $t$$\geq$100 micro-second (corresponding to the applied ac frequency of $\leq$10$^{4}$ Hz).

\begin{figure*}
\centering
\includegraphics[width=15 cm]{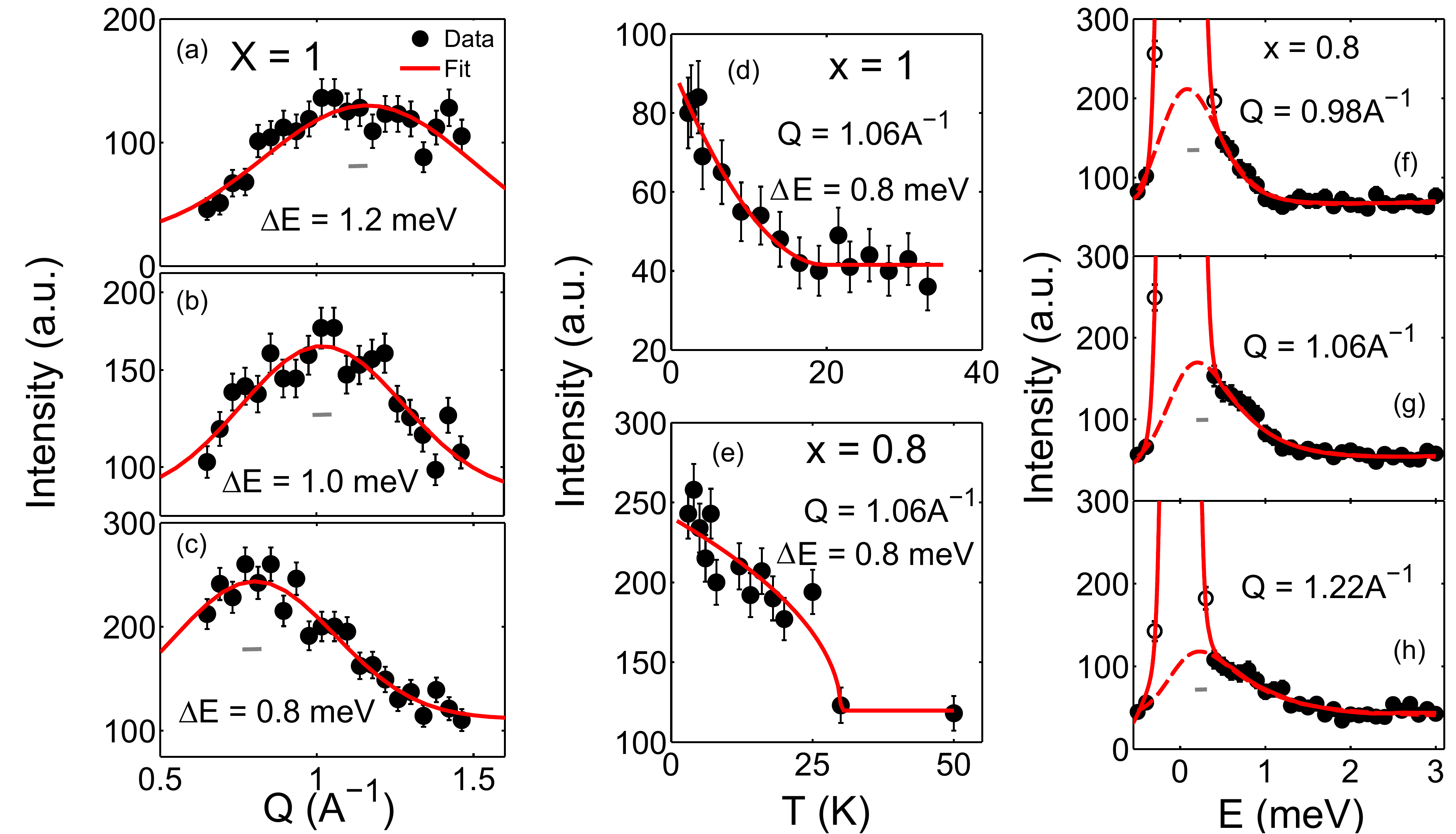} \vspace{-4mm}
\caption{\textbf{Magnetic fluctuations at shorter time scale, as probed by neutron scattering measurements}. (a)-(c) Characteristic Q-scans at constant energies in x=1 composition of Ca(Ir$_{1-x}$Ru$_{x}$)O$_{3}$. Measurements were performed at a cold triple-axis spectrometer with fixed final energy of 3.7 meV, resulting in the energy resolution of $\simeq$ 0.16 meV. The Q-scans are well-described by a Gaussian line shape. Full width at half maximum (FWHM) of the fitted Gaussians are much broader than the instrument spatial resolution (indicated by the grey bar at the center), thus suggesting short-range dynamic structure factor in the system. (d)-(e) The inelastic intensity at $\Delta$E = 0.8 meV is found to be increasing significantly above the background below $T$$_{0}$ = 22 K in x = 1 and $T$$_{0}$ = 28 K in x = 0.8 compositions. Solid red lines are fit to the power law ($I$$\propto$(1-$T$/$T$$_{0}$)$^{-2\beta}$), used to determine the dynamic magnetic transition temperature $T$$_{0}$. (f)-(h) Magnetic fluctuations at short time scale (10$^{-12}$ s) of paramagnetic origin are also evident in x = 0.8 composition. Broad quasi-elastic peaks are described by a Lorentzian curve convoluted with the Gaussian instrument resolution. The peak intensity becomes weaker at higher Q (following the magnetic form factor dependence on Q). Error bar represents one standard deviation in all figures.
} \vspace{-4mm}
\end{figure*}

\textbf{Investigation of magnetic fluctuations at short time scale}. Magnetic fluctuations on the shorter time scale are probed using inelastic neutron scattering measurements in two stoichiometries of x = 1 and 0.8. These compositions were chosen for two reasons: first, less iridium content in these compounds reduces absorption of neutrons and second, magnetic fluctuations tend to be stronger for x$\geq$ 0.3. Inelastic neutron scattering measurements were performed on a cold triple axis spectrometer with an energy resolution (FWHM) of 0.16 meV (see Methods for detail). Representative scans in Q at constant energy transfers $\Delta$E (= E$_{i}$ - E$_{f}$), varying between 0.8-1.5 meV, and in energy at constant Q's are plotted in Fig. 4. The Q-scans at finite energy transfers represent the dynamic structure factor of correlated spins, fluctuating on a time-scale limited by the instrument resolution of $\simeq$10$^{-12}$ seconds. The background corrected data in Fig. 4a-c (corresponding to x = 1 composition) are well described by Gaussian lineshape of width much bigger than the instrument Q-resolution, demonstrating the short-range dynamic order in the system. The spatial extent of the short-range dynamic order extends almost up to two unit cells along the $c$-axis or equivalently almost three unit cells along $a$ or $b$-axis. Further information about the spin fluctuations in this composition is obtained from the temperature dependent study of the dynamic structure factor. As shown in Fig. 4d, the inelastic intensity of the short-range dynamic structure factor at $\Delta$E = 0.8 meV exhibits a subtle rise below $T\leq 22$ K. 

Similar dynamic properties are observed in the Ir-doped compound of x = 0.8 composition. As illustrated in Fig. 4f-4h, the background corrected data is well described by a Lorentzian lineshape convoluted with the Gaussian instrument resolution. The typical line-width ($\Gamma$, full width at half maximum) of the Lorentzian curve, $\Gamma$$\simeq$0.5 meV, is found to be much broader than the spectrometer's resolution ($\simeq$0.16 meV). The integrated intensity of the quasi-elastic peak decreases at higher Q, following the magnetic form factor; hence, confirming magnetic nature of the fluctuations at a time scale limited by the instrument resolution $\simeq$ pico-seconds. We have also performed a temperature dependent study of the dynamic structure factor, which shows that the inelastic intensity at $\Delta$E = 0.8 meV increases significantly above the background below $T$$\simeq$28 K (Fig. 4e). The onset of the dynamic magnetism takes place at a higher temperature in x = 0.8 composition compared to x = 1 composition. This observation is consistent with the dynamic susceptibility measurements where strong magnetic fluctuations, as depicted by temperature and frequency dependences of $\chi$$^{"}$ in Fig. 3, persist to higher temperature in x = 0.8 composition, compared to x = 1. In both cases, however, magnetic ions are found to fluctuate at both the short and the long time scales. Similar behavior of spin fluctuations at the short time scale is expected to occur in other compositions with x $\geq$ 0.3, where strong magnetic fluctuations at the long time scale are detected in the dynamic susceptibility measurements. Finally, we emphasize that the same regime of chemical doping (0.3$\leq$x$\leq$1.0) is associated to the metallic phases in Ca(Ir$_{1-x}$Ru$_{x}$)O$_{3}$. Neutron scattering measurements were also performed in applied magnetic field to explore possible effects on the static and the dynamic magnetic properties. As shown in Fig. S2 (in Supplementary Materials), no change is observed between the elastic patterns obtained at $H$ = 0 T ($T$ = 1.5  K) and $H$ = 10 T. Inelastic measurements also manifest similar behavior in both x = 1 and x = 0.8 compositions (see Fig. S3 and Fig. S4 in Supplementary Materials). 

\begin{figure}
\centering
\includegraphics[width=9 cm]{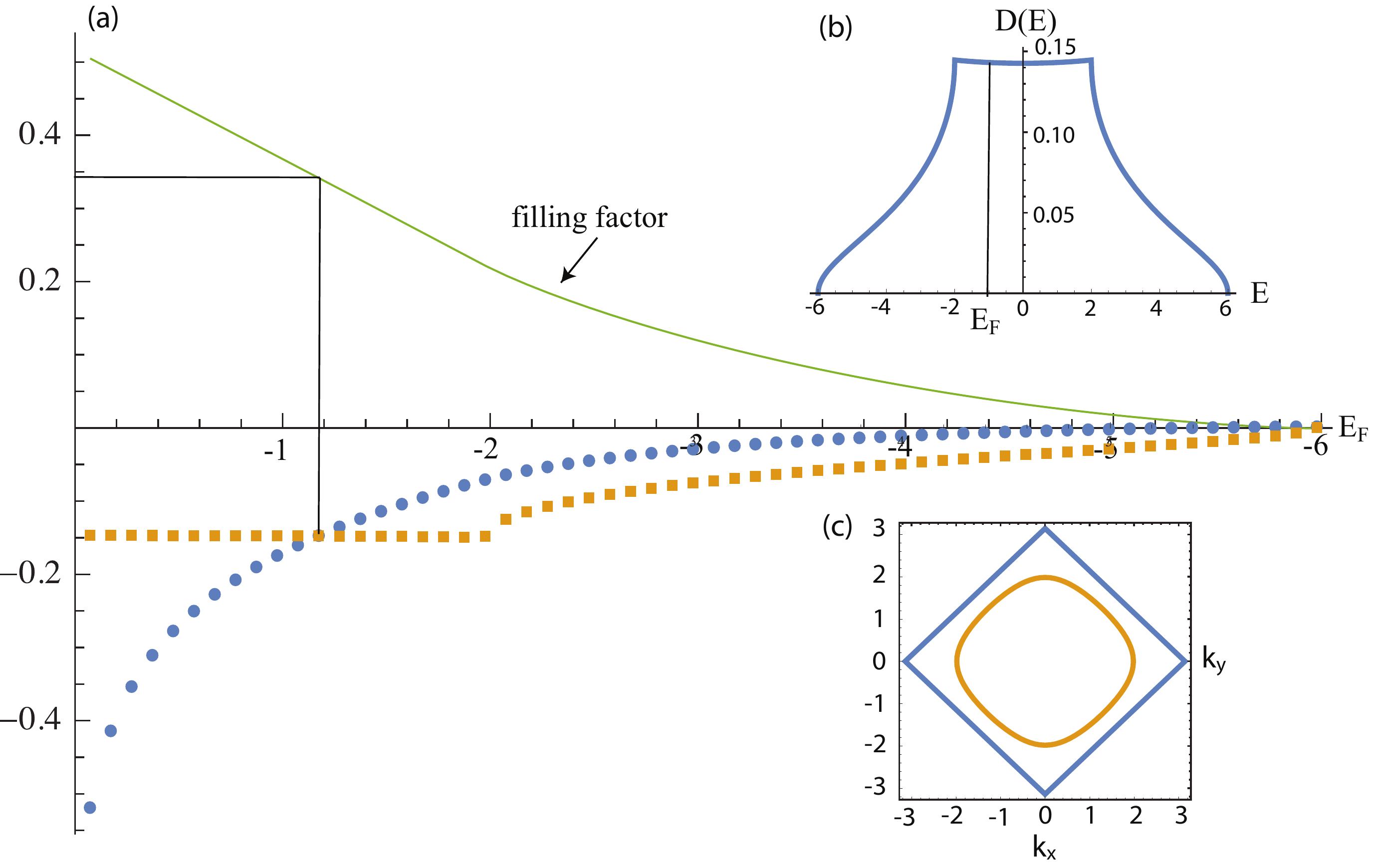} \vspace{-4mm}
\caption{\textbf{Theoretical estimation of the insulator-to-metal cross-over properties in a three dimensional Hubbard band}. (a) Plots of the static ferromagnetic spin susceptibility $\chi({\bf 0})$ (square dots), the antiferromagnetic spin susceptibility $\chi(\pi,\pi)$ (circular dots),  and the filling factor (solid line), versus Fermi energy, $E_F$,  for a three-dimensional Hubbard band (dispersion $\epsilon_{\bf k}=-2t [\cos (k_xa)+\cos (k_ya)+\cos (k_za)]$, where ${\bf k}=(k_x,k_y,k_z)$ is the electron wave vector, $t$ is the hopping amplitude, and $a$ is the lattice constant). Both $t$ and $a$ are taken to be $1$ in this plot. Observe how the ferromagnetic susceptibility becomes larger than the antiferromagnetic susceptibility at a filling factor  $< 0.35$, which is roughly 30\% lower than half filling, corresponding to a doping level of about 0.3  (b) Density of states of the three dimensional Hubbard model vs energy  (c) Fermi surface of the three dimensional Hubbard model in the $k_z=\pi/2$ plane at half-filling (outer square) and 35\% filling (inner loop).  The reduction of nesting, which promotes anti ferromagnetism near half filling, is responsible for the crossing from predominantly antiferromagnetic to predominantly ferromagnetic correlations.} 
\label{PhaseDiagram}
\end{figure}

\textbf{Antiferromagnetic to ferromagnetic crossover in a three dimensional Hubbard band}. At x=0 the pristine CaIrO$_3$ is known to be an antiferromagnetic Mott insulator.  This system has a half-filled d-band  arising from d-orbitals of Ir (configuration $d^5$), which are first split by crystal fields into six $t_{2g}$ and  four $e_g$ states, and then further split by strong spin-orbit interaction, within the $t_{2g}$ multiplet, into a quartet and a doublet of states -- the latter being the half-filled band, which one can roughly model as a half-filled Hubbard band.\cite{Imada} The ground state of this model is an antiferromagnetic insulator.  Chemical substitution  of the lighter element Ru, with one electron less than Ir in the d band (configuration $d^4$) and much weaker spin-orbit interaction, is expected to destroy the simple one-band Hubbard model.  If, however, the doping level is sufficiently low, we can still assume that the model retains, at least qualitatively, its validity.  Under this assumption we can view the Ru substitution as a form of  ``hole doping" in the Hubbard model: in the simplest mean field theory, this is expected to lead to a ferromagnetic metallic phase.\cite{Nagaoka,Zaanen} Quantum fluctuations, beyond mean field theory, are expected to wash out long-range ferromagnetism, while preserving strong ferromagnetic fluctuations and, of course, the metallic character of the state. Experimentally, this is precisely what we observe, with the metal insulator transition occurring at the relatively small value of doping, x=0.3, and a concomitant change in the character of the magnetic fluctuations from static antiferromagnetic to dynamical ferromagnetic. Remarkably, the value x=0.3 coincides with the level of hole doping for which the wave vector dependent magnetic susceptibility of the Hubbard model changes from being predominantly antiferromagnetic (peak at wave vector ${\bf q}=(\pi,\pi)$) to be predominantly ferromagnetic (peak at ${\bf q}=0$). The situation is described in Fig.~\ref{PhaseDiagram} and its caption.

\textbf{Discussion}

We have investigated the role of dynamic magnetism in the metal-insulator transition process in the family of strongly correlated  perovskite compounds Ca(Ir$_{1-x}$Ru$_{x}$)O$_{3}$. We have shown that the insulator-to-metal transition is accompanied by unambiguous magnetic fluctuations as a function of the chemical doping coefficient x. As the chemical doping coefficient x increases, the system crosses over into the metallic regime with the concurrent development of strong dynamic magnetism. It looks as if the static magnetic order, found in low x compositions, had melted away at x $\simeq$ 0.3 and the magnetic ions, ``emboldened" by the energy obtained from this melting process, suddenly become highly dynamic. Hence, it is no surprise that the strongest dynamic susceptibility, depicting magnetic fluctuations, is observed for x = 0.3 chemical composition.  This is the same composition (x = 0.3) where the onset to the metallic behavior is detected. Therefore, a direct connection between the insulator-to-metal transition and magnetic fluctuations is quite unequivocal. A theoretical analysis of the possible underlying mechanism behind this process is consistent with the experimental evidence. 

Magnetic fluctuations and the associated quantum magnetic critical behavior are often considered to be the underlying physics behind the unconventional superconductivity in magnetic materials of the strongly correlated electrons origin.\cite{Tranquada,McLaughlin,Fisk,Pengcheng} However, a material first becomes a metal, in general, before exhibiting superconductivity as a function of temperature. Therefore, the same magnetic fluctuations that are considered to be at the core of the unconventional superconductivity -- such as cuprates, pnictides or the heavy electron systems --  must play a key role in driving the metallic behavior. Despite the heavy interest in this topic, the situation is still not well understood. Our joint experimental and theoretical  effort  provides new information about the insulator-to-metal transition in strongly correlated electronic systems, which in turn will help us develop a robust framework for understanding unconventional superconductivity in these materials.

\textbf{Methods}

The high purity polycrystalline samples of Ca(Ir$_{1-x}$Ru$_{x}$)O$_{3}$ were synthesized by conventional solid-state reaction method using ultra-pure ingredients of IrO$_{2}$, RuO$_{2}$ and CaCO$_{3}$. Starting materials were mixed in stoichiometric composition, with five percent extra RuO$_{2}$ to compensate for their rapid evaporation (in Ru-doped perovskites), palletized and sintered at 950$^{o}$ for three days. The furnace cooled samples were grinded, palletized and sintered at 1000$^{o}$ for another three days. Resulting samples were characterized using Siemens D500 powder X-ray diffractometer,\cite{NIST} confirming the single phase of material. The X-ray diffraction data were analyzed using a widely used commercial software JADE.\cite{NIST} As shown in Fig. S1 of the supplementary materials, every single peak of the XRD pattern is identified with the orthorhombic structure of Ca(IrRu)O$_{3}$. Four probe technique was employed to measure electrical properties of Ca(Ir$_{1-x}$Ru$_{x}$)O$_{3}$ using a closed-cycle refrigerator cooled 9 T magnet with measurement temperature range of 1.5-300 K. Detailed ac susceptibility measurements were performed using a Quantum Design Physical Properties Measurement System with a temperature range of 2-300 K.\cite{NIST} Neutron scattering measurements were performed on the pristine powder samples of 3.9 g of CaRuO$_{3}$ and 3.2 g of CaIr$_{0.2}$Ru$_{0.8}$O$_{3}$ on the spin-polarized triple-axis spectrometer at the NIST Center for Neutron Research with fixed final neutron energy of 3.7 meV. Neutron Scattering measurements employed a cold BeO-filter followed by a radial collimator and the focused analyzer. The incoherent scattering from vanadium is used to determine the spectrometer's resolution in the specified configuration. At this fixed final energy, the spectromete's resolution (FWHM) was determined to be $\simeq$ 0.16 meV.

\section{Acknowledgements}

This work used facilities supported in part by the Department of Commerce. DKS acknowledges support from the MU IGERT program, funded by NSF under grant number DGE-1069091.

\section{Author contributions}

D.K.S envisaged the research idea. Samples were synthesized by J.G. and A.D. The measurements and experimental data analysis were carried out by J.G., L.H., T.H., A.D. and D.K.S. Theoretical analysis was carried out by G.V. and the paper was written by D.K.S. and G.V.

\section{Additional information}

\textbf{Competing Financial Interests} Authors declare no competing financial interests.

\clearpage


\begin{thebibliography}{99}
\bibitem{NIST} National Institute of Standards and Technology or the Department of Commerce does not endorse any particular commercial product.
\end{thebibliography}
\end{document}